\documentclass[
    ,final            
  ]
  {aipproc}

\layoutstyle{6x9}

\begin{document}

\author{John Ellis}{
  address={Theoretical Physics Division, Physics Department, CERN, CH 1211 Geneva 23, Switzerland},
  email={John.Ellis@cern.ch}
}


\title{Beyond the Standard Model \\
at the LHC and Beyond}


\begin{abstract}
  Many of the open questions beyond the Standard Model will be addressed by
  the LHC, including the origin of mass,
  supersymmetry, dark matter and the possibility of large extra dimensions.
  A linear $e^+ e^-$ collider (LC) with sufficient  centre-of-mass energy would
  add considerable value to the capabilities of the LHC.
\end{abstract}

\maketitle


\begin{center}
CERN-PH-TH/2007-174
\end{center} 

\bigskip

\section{Open Questions beyond the Standard Model}

There is a standard list of fundamental open questions beyond the Standard Model:\\
$\bullet$ What is the origin of particle masses?\\
Are they indeed due to a Higgs boson, as hypothesized within the Standard Model? 
If so, is the Higgs boson accompanied by some other physics such as
supersymmetry? If not, what replaces the
Higgs boson?\\
$\bullet$ Why are there so many types of matter particles?\\
Related to this question is the mixing of the different flavours of quarks and leptons, and 
the mechanism for CP violation. This matter-antimatter difference is thought to be
linked to cosmological baryogenesis. 
However, the Standard Model cannot, by itself,
explain the cosmological matter-antimatter asymmetry, adding urgency to the search
for flavour and CP violation beyond the Standard Model.\\
$\bullet$ Are the fundamental forces unified?\\
If so, in the simplest models this unification occurs only at some very high energy 
$\sim 10^{16}$~GeV. Physics at this scale may be
probed via neutrino physics, or possibly less directly at accelerators
via measurements of particle masses and couplings, and looking
for unification relations between them.\\
$\bullet$ What is the quantum theory of gravity?\\
The best candidate for such a theory may be (super)string theory,
which generously predicts extra space-time dimensions as well as supersymmetry, but at
what energy scale?

The good news is that all of these fundamental open questions will be addressed by
the LHC: its energy should be ample for resolving the problem of mass, including the questions
whether there is a Higgs boson and/or supersymmetry, a dedicated
experiment will be examining matter-antimatter differences, models of unification could
be probed via measurements of sparticle masses and couplings, and string theory
might be probed via supersymmetry breaking, extra dimensions or even black hole
production and decay. Accordingly, most of this talk will be concerned with LHC
physics, accompanied by some asides about linear collider (LC) physics.

Supersymmetry may play a role in answering
most of the open questions, for example by stabilizing the scale of electroweak
symmetry breaking or by aiding the unification of the gauge coupling, and it also seems
to play an essential role in string theory. Therefore, I have invested many or my personal
efforts in supersymmetry, and apologize for giving
pride of place to it in the rest of this talk.

\section{Higgs Physics}

\begin{figure}
\resizebox{1.0\columnwidth}{!}
{\includegraphics{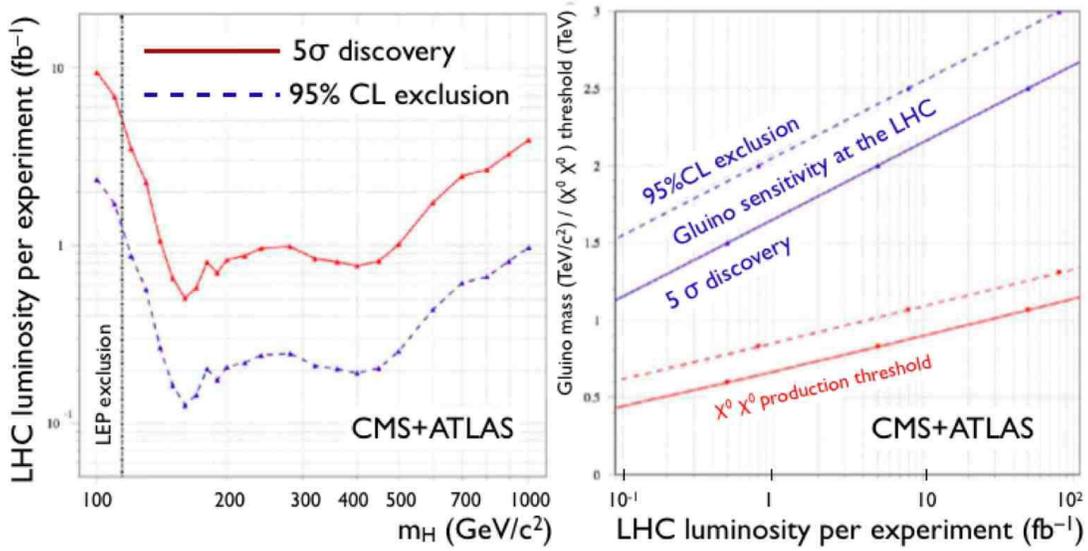}}
\caption{\label{fig:Higgs} The combined sensitivities of ATLAS and CMS to a
Standard Model Higgs boson (left), and the gluino (right), as a function of the
analyzed LHC luminosity. The right panel also shows the threshold for
sparticle pair production at a LC for the corresponding gluino mass, calculated
within the CMSSM~\protect\cite{POFPA}.}
\end{figure}

Several different Higgs production mechanisms will be important at the LHC,
including gluon-gluon and $WW$ fusion, and production in association with
gauge bosons or heavy quarks. Several decay modes will also be important, including
$\gamma \gamma$, four leptons, $\tau^+ \tau^-$, $WW$, ${\bar b}b$, etc.~\cite{ATLAS,CMS}.
Accordingly the search for the Higgs boson at the LHC will require the combination of
many different signatures and hence excellent understanding of all components of 
ATLAS and CMS. As seen in Fig.~\ref{fig:Higgs},
once both detectors have accumulated and analyzed a couple of
hundred pb$^{-1}$ of data, they may be able to start excluding certain ranges of Higgs
masses at the 95~\% confidence level~\cite{POFPA}. 
With $\sim 1$~fb$^{-1}$ of analyzed data, they might
be able to exclude a Standard Model Higgs boson over the entire mass range up to 1~TeV, or
they might be able to establish a five-$\sigma$ signal if the Higgs boson weighs between 
$\sim 150$ and 500~GeV. Several fb$^{-1}$, corresponding to several months at one
tenth of the design luminosity, would be needed to discover the Higgs boson over the
entire mass range.

With some luck, it may also be possible to determine the spin of the Higgs 
boson~\cite{Choi,Buszello,Accomando}.
For example, if it is light and is seen to decay into $\gamma \gamma$, it cannot have
spin one. If it has higher mass and is seen to decay into four leptons, their angular
correlations will enable $J^P$ states other than $0^+$ to be excluded with high
significance. Analogous studies could be made at a LC, also using angular
distibutions and the threshold behaviour for $e^+ e^- \to Z H$~\cite{LCspin}.

\begin{figure}
\resizebox{0.6\columnwidth}{!}
{\includegraphics{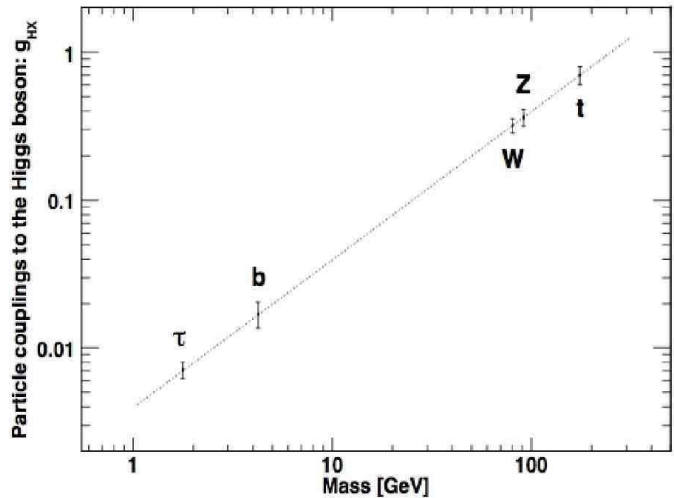}}
\caption{\label{fig:couplings} Estimates of the accuracy with which 
experiments at the LHC could measure the couplings of the Higgs boson to various
particles~\protect\cite{POFPA}.}
\end{figure}

As seen in Fig.~\ref{fig:couplings},
the LHC may also make a first analysis of the proportionality of the Higgs couplings
to particle masses~\cite{POFPA}. 
If $m_H \sim 120$~GeV, it may be able to measure the couplings 
to $\tau^+ \tau^-$, ${\bar b}b$, $WW$, $ZZ$ and ${\bar t}t$ each with an accuracy
$\sim 20$~\%. Much more precise measurements could be made subsequently with
a LC~\cite{LCcouplings}.

As well as its importance for particle physics, the discovery of a Higgs boson
would also be very important for cosmology. An elementary Higgs boson would have caused a
phase transition in the early Universe when it was $\sim 10^{-12}$~s old, and might
have generated the matter in the Universe via electroweak
barygenesis. Further back in the history of the Universe, a related inflaton might
have expanded the Universe exponentially when it was $\sim 10^{-35}$~s old.
Coming back to the present, naively the Higgs boson of the Standard Model
would contribute a factor $\sim 10^{56}$ too much to the present-day dark energy,
apparently requiring some `miraculously' fine-tuned cancellation. Cosmologists
should be as interested as particle physicists in the d\'enouement of the Higgs saga.
For the time being, the LHC and a possible subsequent LC will be our only direct
windows on this physics.

\section{Supersymmetry?}

My personal favourite candidate for new physics beyond the Higgs boson is supersymmetry,
for several reasons. True, it is intrinsically beautiful and (almost) an essential ingredient in string
theory, etc., but these are not the reasons that motivate me to expect that it may appear at the
LHC. There are four specific reasons why one might expect 
supersymmetry to appear around the TeV scale, and hence be accessible to the LHC.
One is the naturalness or hierarchy problem~\cite{hierarchy}, another is the unification of the gauge
couplings~\cite{GUTs}, another is the supersymmetric prediction of a light Higgs boson as
preferred by the precision electroweak data~\cite{Higgs}, and another is that many supersymmetric
models predict the existence of cold dark matter with a density comparable to that
required by astrophysics and cosmology~\cite{EHNOS}.


In the following, I concentrate on the minimal supersymmetric extension of the
Standard Model (MSSM). In addition to the coupling $\mu$ between its two gauge
multiplets, and the ratio $\tan \beta$ of their vacuum expectation values, the MSSM
has many apparently arbitrary soft supersymmetry-breaking parameters, including
scalar masses $m_0$, gaugino masses $m_{1/2}$, trilinear soft couplings $A_0$ and
a bilinear soft coupling $B_0$. It is commonly assumed that the $m_0, m_{1/2}$ and $A_0$
are each universal at some input GUT scale, a framework known as the constrained MSSM
(CMSSM). This is not the same as minimal supergravity (mSUGRA), which fixes in addition
the gravitino mass: $m_{3/2} = m_0$, and imposes $B_0 = A_0 - m_0$. We will see later
some potential implications of these extra conditions.

There are direct limits on sparticle masses from their absences at  LEP and the Tevatron,
and indirect constraints from the LEP lower limit $m_h > 114$~GeV and from $B$
physics, including in particular measurements of $b \to s \gamma$ decay. One
possible indication of new physics at the TeV scale may be provided by the BNL
measurement of the anomalous magnetic moment of the muon~\cite{g-2}, that seems to
exhibit a three-$\sigma$ discrepancy with the Standard Model, though this is still
somewhat controversial. The strongest constraint on (one combination of)
supersymmetric model parameters is provided by the density of cold dark matter:
$0.094 < \Omega_\chi h^2 < 0.124$, assuming that it is mainly composed of the
lightest neutralino $\chi$. 

This not the only possibility: presumably the LSP should
have neither strong nor electromagnetic interactions, but there are other
candidates that also have these properties. The supersymmetric partners of the neutrinos have
been excluded by a combination of LEP and direct dark matter searches, but the LSP
might be the spartner of some particle beyond the Standard Model, such as the gravitino.

\begin{figure}
\resizebox{0.45\columnwidth}{!}
{\includegraphics{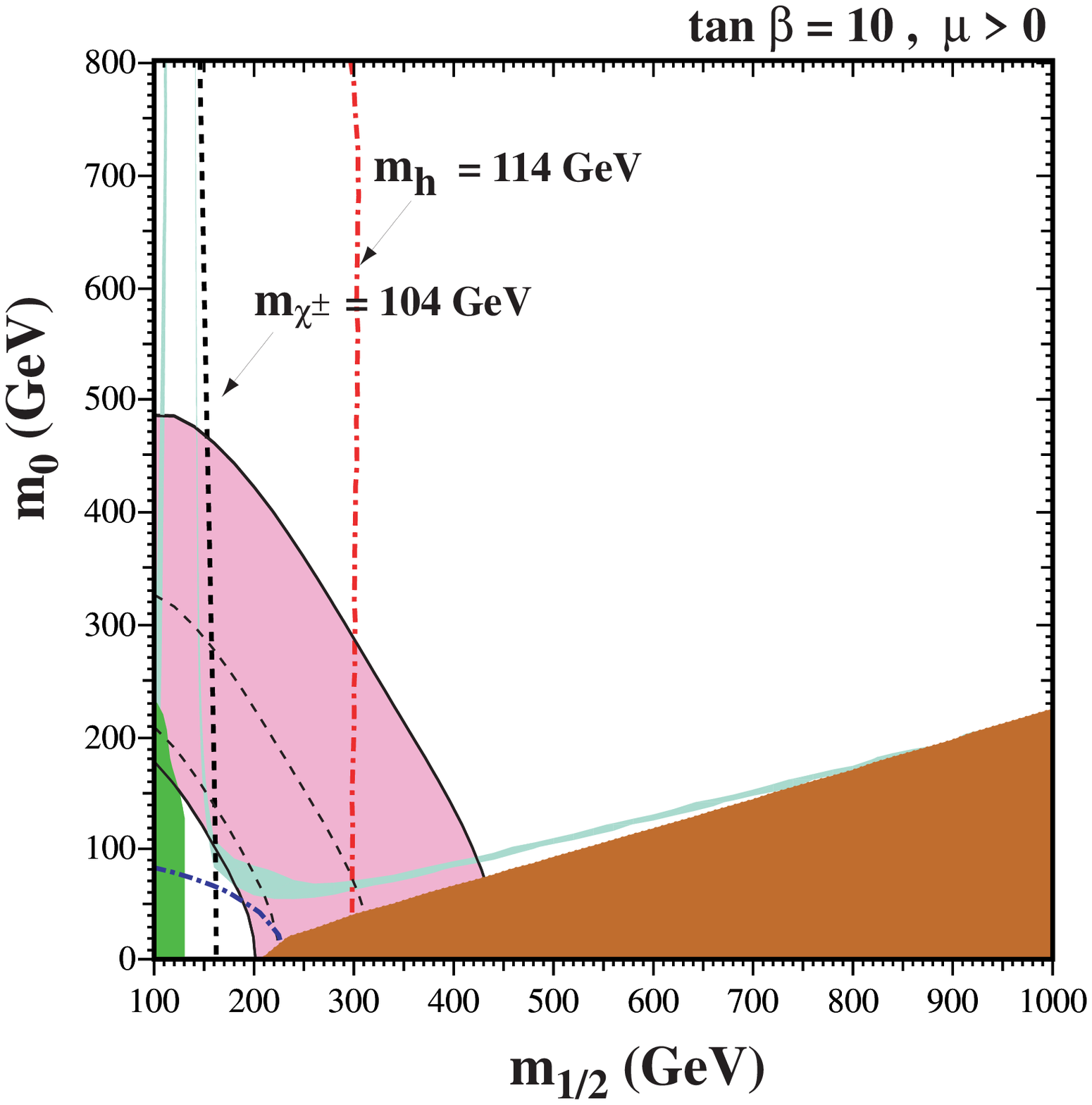}}
\resizebox{0.55\columnwidth}{!}
{\includegraphics{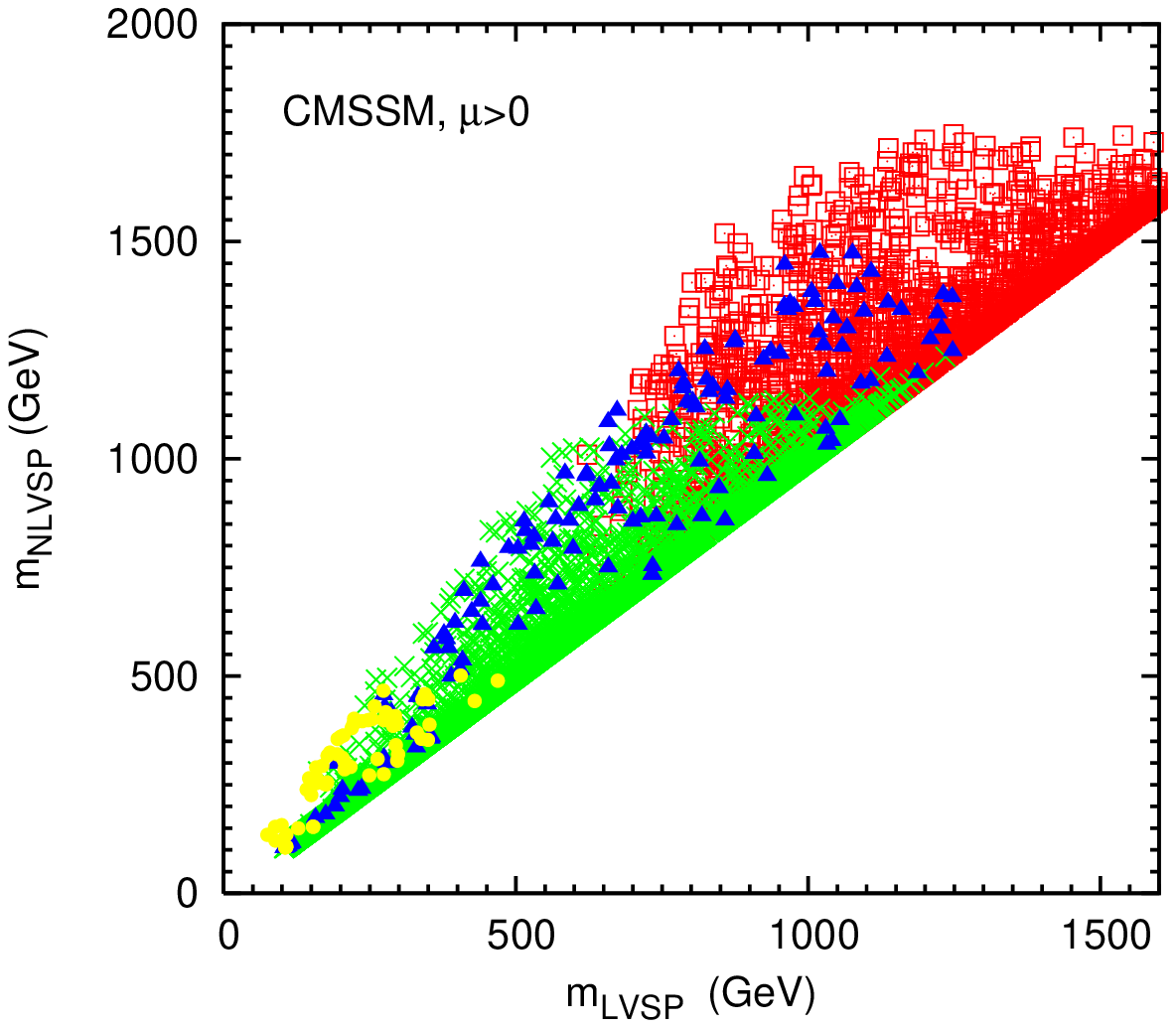}}
\caption{\label{fig:CMSSM} The $(m_{1/2}, m_0)$ plane in the CMSSM for
$\tan \beta = 10$, $\mu > 0$ and $A_0 = 0$, incorporating the theoretical, experimental
and cosmological constraints described in the text (left), and the masses of the
lightest and next-to-lightest visible supersymmetric particles in a sampling
of CMSSM scenarios (right)~\protect\cite{EOSS}. The right panel also indicates the scenarios
providing a suitable amount of cold dark matter (blue), those detectable at the LHC
(green) and those where the astrophysical dark matter might be detected
directly (yellow).}
\end{figure}

In a minimal supersymmetric model with universal soft supersymmetry-breaking
parameters, one may consider these constraints in the $(m_{1/2}, m_0)$ plane. As shown
in the left panel of 
Fig.~\ref{fig:CMSSM}, where the LSP is assumed to be the lightest neutralino, the resulting
allowed regions include a narrow (pale turquoise)
strip near the boundary beyond which $m_\chi > m_{\tilde \tau_1}$ (dark brown shading)
where $m_0 \sim 100$~GeV, and another strip (not shown) where $m_0 > 1$~TeV near the
boundary beyond which
electroweak symmetry breaking is no longer possible (this region is disfavoured by $g_\mu - 2$,
indicated by pink shading).

The fact that these allowed strips extend to relatively large values of $m_{1/2}$ and $m_0$
implies that even the lightest visible sparticles may not be very light, as seen in
the right panel of
Fig.~\ref{fig:CMSSM}. We generated a large sample of possible CMSSM scenarios
(the red symbols), of which a subsample (shown in darker blue) provide the right amount
of cold dark matter~\cite{EOSS}. Also shown (in paler green) are the scenarios that would be
detectable at the LHC. We see that most (but not all) of the dark matter scenarios
should be visible at the LHC: one may be hopeful, but there is no guarantee that
the LHC will discover supersymmetry. On the other hand, direct
astrophysical dark matter search experiments (very pale yellow) may have less chances in the
foreseeable future. An LC with centre-of-mass energy $\sim 3$~TeV would be needed
to see sparticles in all the CMSSM dark matter scenarios we sampled.


Can one estimate the scale of supersymmetry on the basis of present data~\cite{EHOWW}?
The precision electroweak measurements of $m_W$ and $\sin^2 \theta_W$
both have some sensitivity to $m_{1/2}$ through radiative corrections, and $m_W$
slightly prefers smaller values of $m_{1/2}$, though this trend is hardly
significant. The agreement of $b \to s \gamma$ with the Standard Model
offers no encouragement to enthusiasts for
light supersymmetry, and other $B$-decay observables such as $B_s \to \mu^+ \mu^-$
and $B_u \to \tau \nu$ do not yet provide much information about the possible
scale of supersymmetry breaking. On the other hand, the disagreement between the
experimental value of $g_\mu - 2$ with the theoretical value calculated in the
Standard Model using low-energy $e^+ e^-$ data could be explained by light
supersymmetry, as shown already in the left panel of Fig.~\ref{fig:CMSSM}.

\begin{figure}
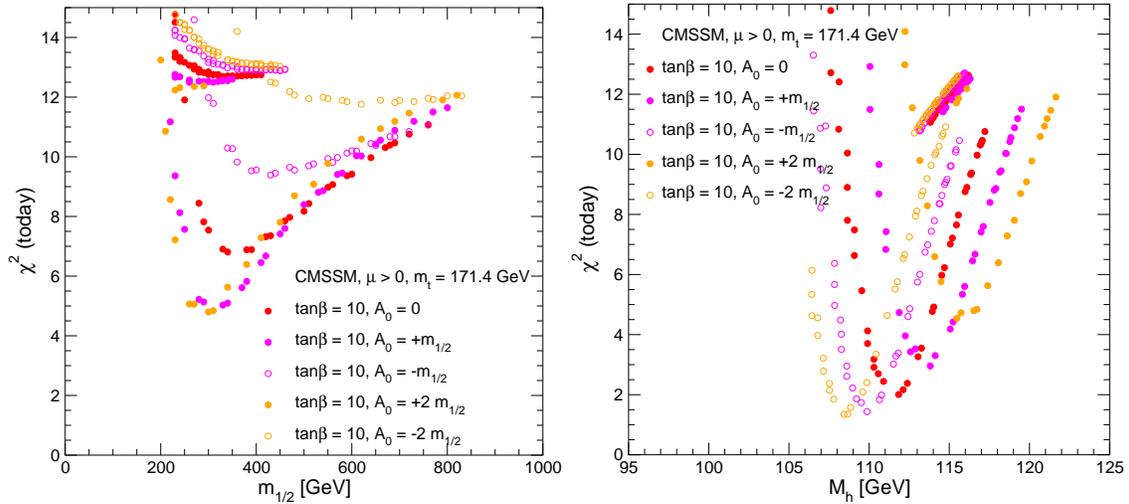

\resizebox{0.5\columnwidth}{!}
{\includegraphics{ehow5.CHI11a.1714.cl.eps}}
\resizebox{0.5\columnwidth}{!}
{\includegraphics{ehow5.Mh13a.1714.cl.eps}}
\caption{\label{fig:EHOWW}Evaluation of $\chi^2$ for a global fit to electroweak
and $B$-decay observables as a function of $m_{1/2}$ (left)
and $m_h$ (right) in the CMSSM for
$\tan \beta = 10$, $\mu > 0$ and different values of $A_0$~\protect\cite{EHOWW}.}
\end{figure}

As shown in the left panel of Fig.~\ref{fig:EHOWW}~\cite{EHOWW},
a global fit to precision electroweak and $B$-decay observables indicates a
preference for relatively small values of $m_{1/2}$. This is due predominantly to
$g_\mu - 2$, but there is some support from the measurements of $m_W$.
Correspondingly, the most likely value for the mass of the lightest supersymmetric 
Higgs boson is only slightly above the LEP lower limit, as seen in the right panel
of Fig.~\ref{fig:EHOWW}.

The classic signature of sparticle production is the appearance of events with
missing energy-momentum carried away by invisible dark matter particles.
Studies indicate that such a signature should be observable at the LHC above
instrumental and Standard Model backgrounds~\cite{ATLAS,CMS}. 
In many supersymmetric scenarios, the LHC would produce a pair of squarks and/or
gluinos primarily, and these would subsequently decay into a cascade of secondary
lighter sparticles, At the end of each decay chain would appear the LSP, namely the
lightest neutralino $\chi$ in the CMSSM framework considered so far. Studies
indicate that in many benchmark scenarios a number of different sparticle species
could be observed at the LHC and their masses measured quite accurately~\cite{HP,Bench}.
A LC would produce democratically all the sparticle species that are kinematically
accessible, and therefore would be particularly interesting for producing sparticles
without strong interactions, such as sleptons, charginos and neutralinos, which are
expected to be lighter than squarks and gluinos in CMSSM scenarios.

Even a limited amount of LHC
luminosity would enable the LHC to see squarks or gluinos weighing a TeV
or more. As shown in the right panel of Fig.~\ref{fig:Higgs}, it is 
estimated~\cite{ATLAS,CMS} that 0.1~fb$^{-1}$
of LHC luminosity would be sufficient to observe a gluino with a mass of 1.2~GeV
at the five-$\sigma$ level, or to exclude a gluino weighing $< 1.5$~TeV. The
discovery and exclusion reaches would extend to about 2.2 and 2.5~TeV, respectively
with 10~fb$^{-1}$ of LHC luminosity.

As also shown in the right panel of Fig.~\ref{fig:Higgs},
assuming universal input sparticle masses at the GUT scale, the corresponding
thresholds for sparticle pair production at a linear $e^+ e^-$ collider would be
0.5 (0.6) or 0.8 (1.0) TeV~\cite{POFPA,EOS3}. Hence, for example, if the LHC discovers the gluino
with 0.1~fb$^{-1}$, one may expect that sparticle pair production would be accessible
to a linear collider with centre-of-mass energy of 0.5~TeV, whereas if the LHC
does not discover the gluino even with 10~fb$^{-1}$, the $e^+ e^-$ sparticle
pair production threshold may be above 1~TeV. At least in such a simple model,
the LHC will tell us how much energy a linear collider would need to find supersymmetry~\cite{EOS3}.

If Nature is kind, and sparticles not only exist but also are quite light, it will be possible
to test directly unification of the gauge couplings and universality of the soft
supersymmetry-breaking scalar masses with high precision, in particular by
comparing measurements at the LHC and the ILC~\cite{ILC}.
However, there is no guarantee that any sparticles will be light enough to be
produced at the ILC, even with a centre-of-mass energy of 1~TeV. For this
reason, and also because even if there are some light sparticles the heavier
ones will be produced only at higher energies, a high-energy LC would be
advantageous, which is why CERN and its collaborators are developing the
CLIC technology that should be capable of reaching 3~TeV in the centre of mass~\cite{CLIC}.


So far, I have concentrated on the CMSSM, assuming that all the soft
supersymmetry-breaking scalar masses are universal at the GUT scale. However,
this may not be the case: in particular, there is no good theoretical or
phenomenological reason why the Higgs scalar masses should be
universal. In models with non-universal Higgs masses (NUHM), there are
two additional degrees of freedom, and the
Higgs mixing parameter $\mu$ and the pseudoscalar Higgs mass $m_A$ may
be treated as free parameters~\cite{NUHM}. 

\begin{figure}
\resizebox{0.5\columnwidth}{!}
{\includegraphics{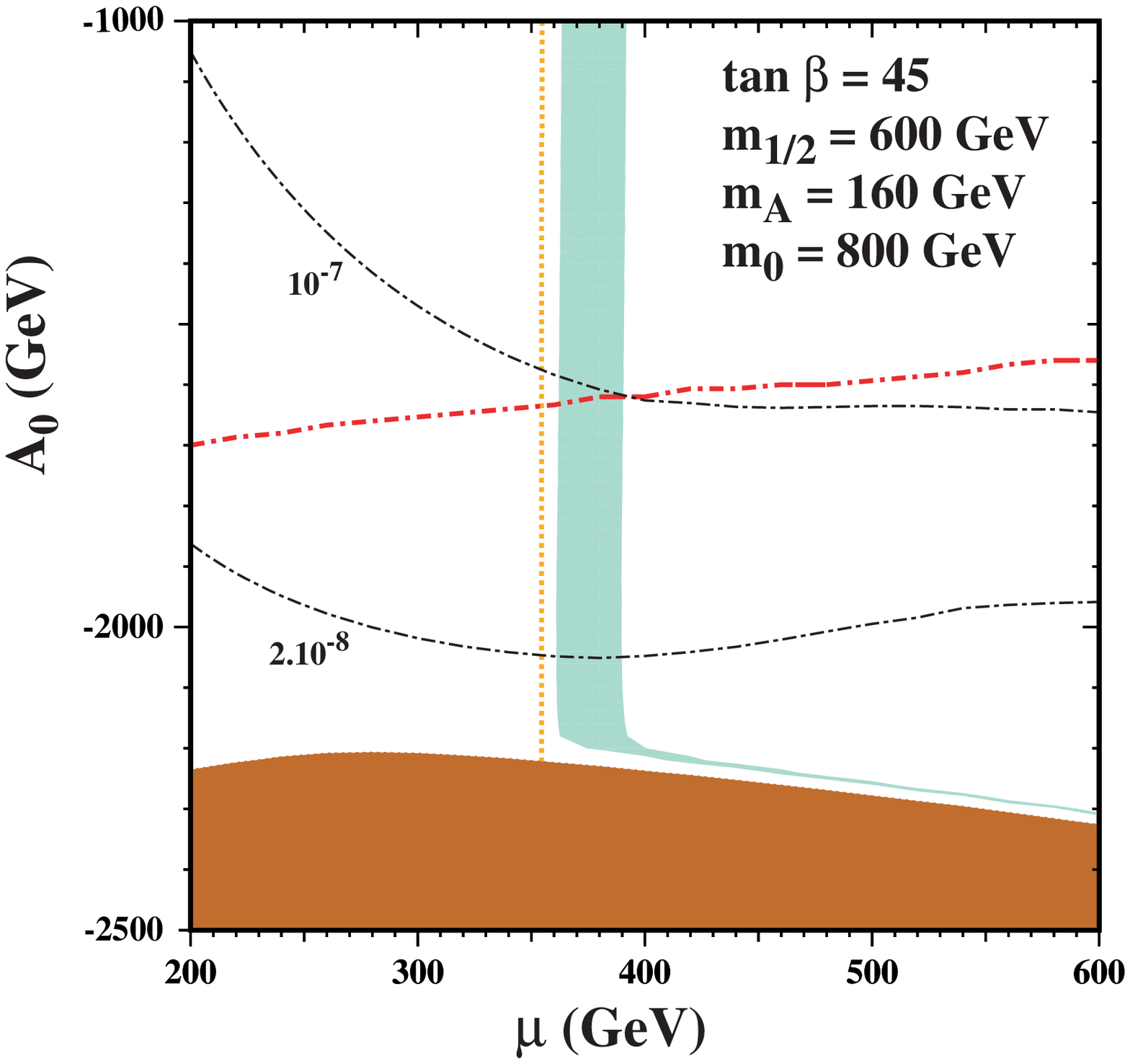}}
\resizebox{0.5\columnwidth}{!}
{\includegraphics{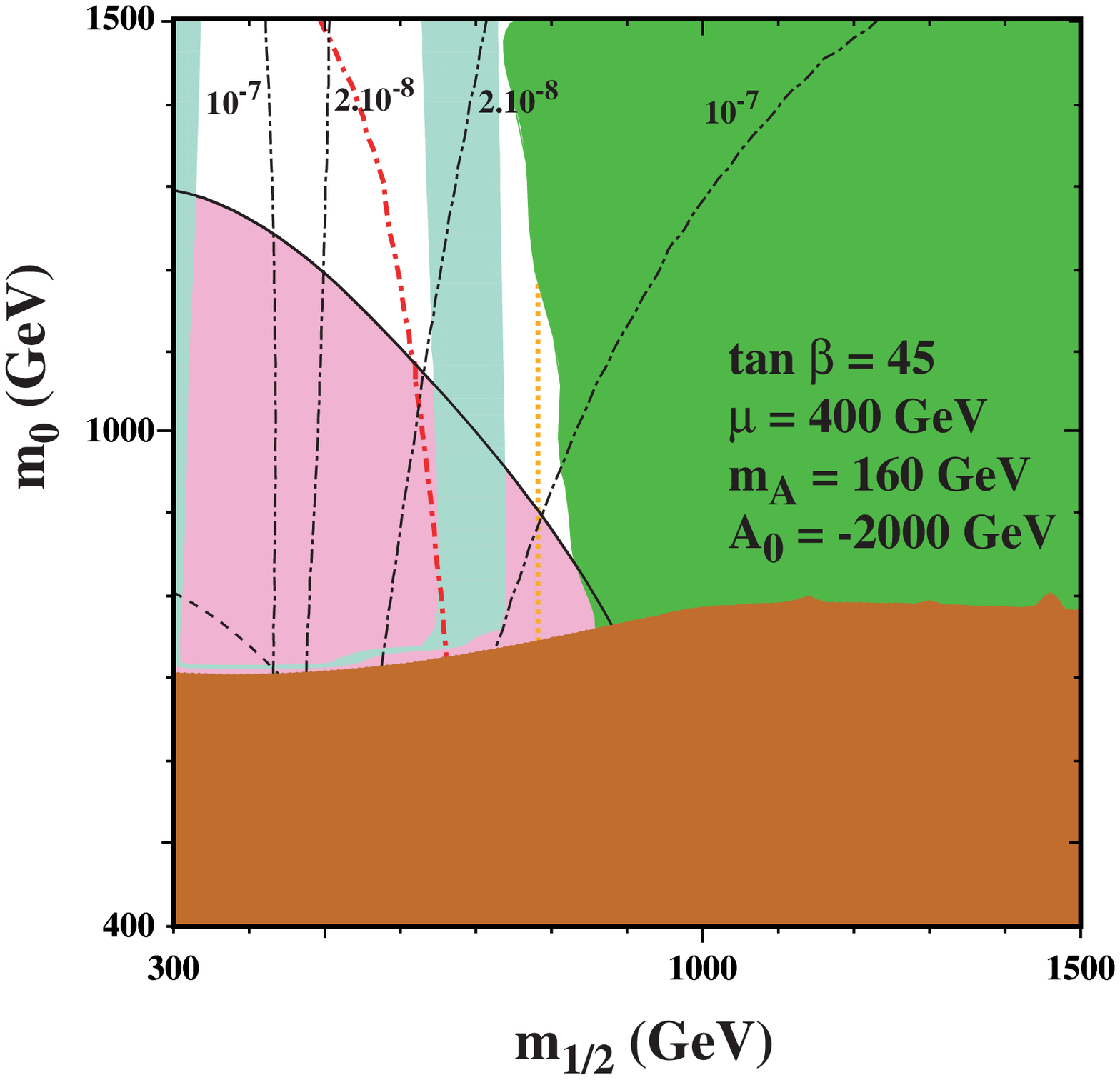}}
\caption{\label{fig:EHOW5}A light $m_A \sim 160$~GeV could be accommodated within
the NUHM, if $\tan \beta \sim 50, m_{1/2} \sim 600$~GeV, $m_0 \sim 800$~GeV,
$\mu \sim 400$~GeV and $A_0 \sim - 2$~TeV, in which case $m_h, b \to s \gamma,
B_s \to \mu^+ \mu^-, B_u \to \tau \nu$ and the cold dark matter scattering
rates would all be very close to the present experimental limits~\protect\cite{EHOW5}. }
\end{figure}

We have studied~\cite{EHOW5} whether the NUHM framework could accommodate a
pseudoscalar supersymmetric Higgs boson as light as the value not yet
excluded by direct searches at the Tevatron. Recall that the CDF
experiment saw a two-$\sigma$ excess in the $\tau^+ \tau^-$ spectrum
that could be explained by supersymmetric Higgs bosons weighing
$\sim 160$~GeV~\cite{CDF}, whereas the D0 experiment saw no such excess~\cite{D0}.
We found~\cite{EHOW5} that $m_A \sim 160$~GeV could indeed be accommodated within
the NUHM, if $\tan \beta \sim 50, m_{1/2} \sim 600$~GeV, $m_0 \sim 800$~GeV,
$\mu \sim 400$~GeV and $A_0 \sim - 2$~TeV, as seen in
Fig.~\ref{fig:EHOW5}. This would be quite an
extreme scenario, but it would have the merit of being testable in the near
future. In this corner of the NUHM parameter space, $m_h, b \to s \gamma,
B_s \to \mu^+ \mu^-, B_u \to \tau \nu$ and the cold dark matter scattering
rates would all be very close to the present experimental limits.

\section{Gravitino Dark Matter?}

The LSP should have no strong or electromagnetic interactions,
otherwise it would bind to conventional matter and be
detectable as some anomalous heavy nucleus.
Apart from the lightest neutralino $\chi$,
possible weakly-interacting scandidates include the sneutrino,
which is excluded by LEP and direct searches in simple models,
and the gravitino ${\tilde G}$, which has attracted relatively little detailed attention
until recently. It would be a nightmare for astrophysical detection, but a
bonanza for the LHC~\cite{Bench3,ERO1,ERO2}, as we now discuss.

\begin{figure}
\resizebox{0.5\columnwidth}{!}
{\includegraphics{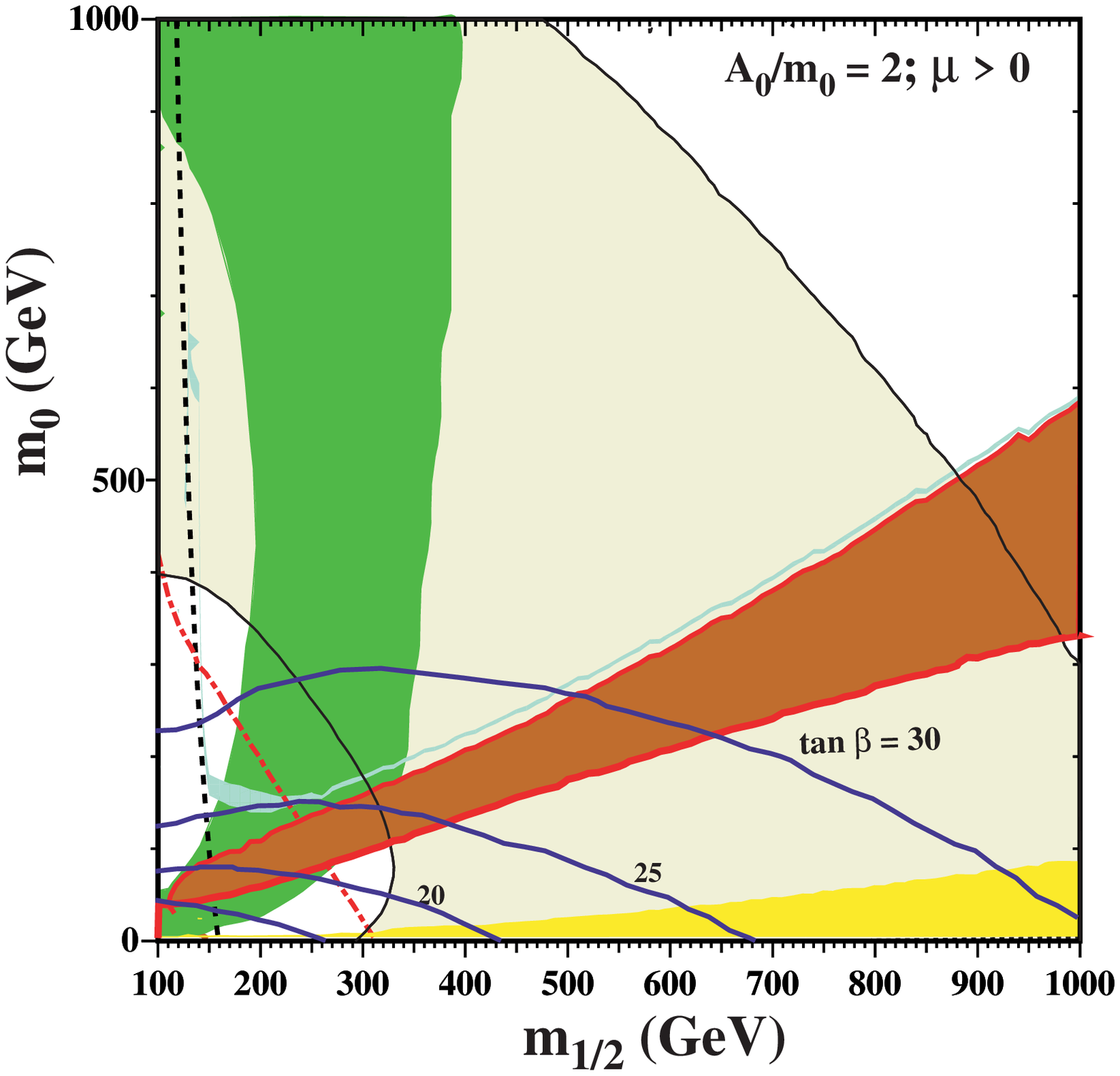}}
\resizebox{0.5\columnwidth}{!}
{\includegraphics{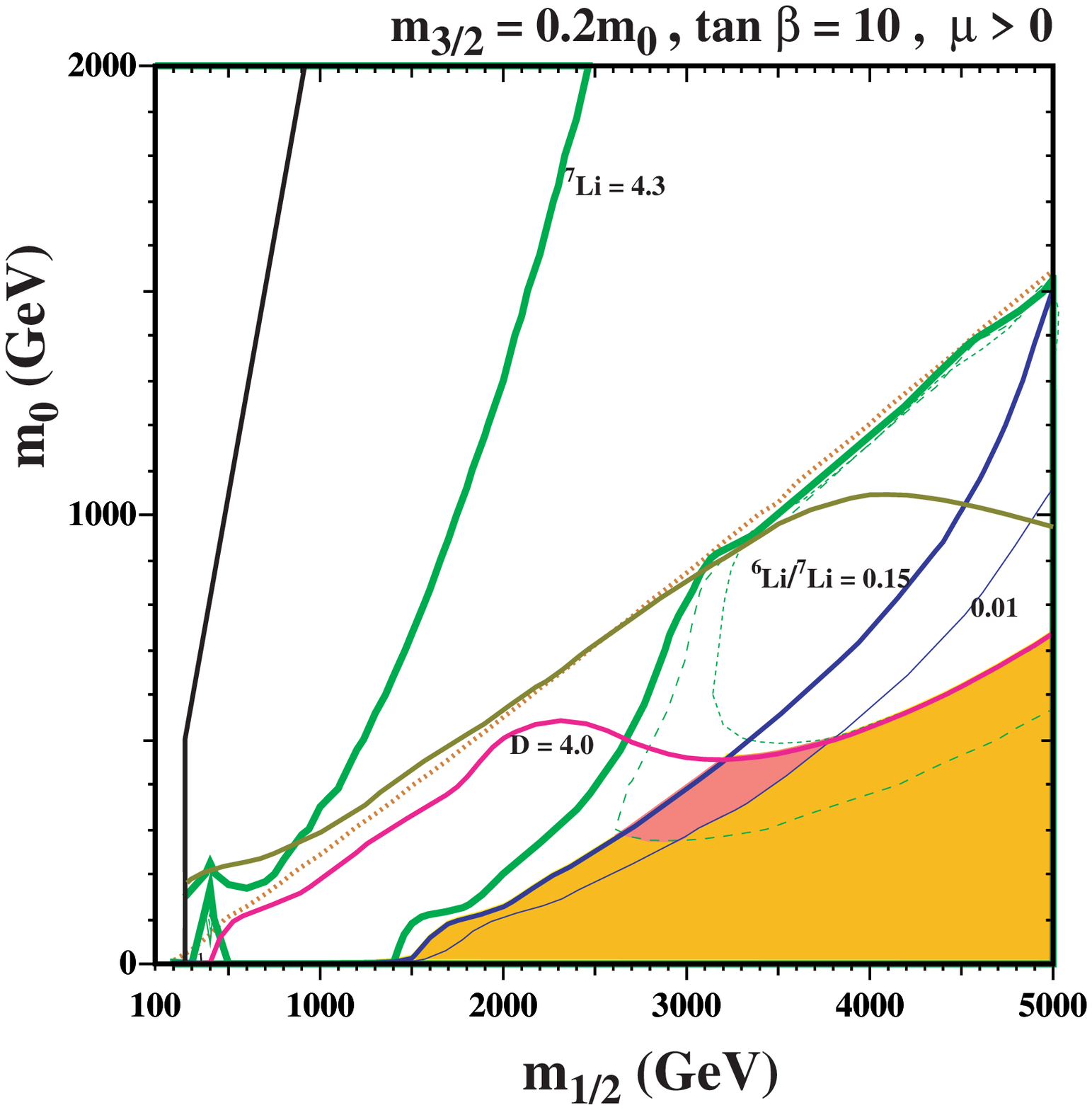}}
\caption{\label{fig:GDM}In the mSUGRA scenario with $A_0 = 2 m_0$ and $\mu > 0$ (left panel),
in addition to the allowed $\chi$ dark matter strip (light turquoise) and the forbidden
${\tilde \tau_1}$ LSP wedge (dark brown), there is a wedge (pale yellow) with a metastable
${\tilde \tau_1}$ whose decays do not disturb the cosmological light element
abundances. In the non-SUGRA scenario with $m_{\tilde G} = 0.2 m_0$ (right panel), there is also
a region (darker pink) where ${\tilde \tau_1}$ bound-state effects~\protect\cite{Pospelov} 
may improve the Lithium abundances~\protect\cite{CEFOS}.}
\end{figure}

The next-to-lightest sparticle (NLSP) would have a very long lifetime in
models with gravity-mediated supersymmetry breaking, due to the weak
gravitational strength of the interactions responsible for its decays. For
example, if the NLSP is the lighter stau slepton ${\tilde \tau_1}$ one has
\begin{equation}
\Gamma_{{\tilde \tau_1} \to \tau {\tilde G}} \; = \; \frac{1}{48 \pi} \frac{1}{M_P^2}
\frac{m_{\tilde \tau_1} ^5}{m_{\tilde G}^2} \left(1 - \frac{m_{\tilde G}^2}{m_{\tilde \tau_1} ^2} \right)^4,
\label{gravrate}
\end{equation}
leading to a lifetime that could be measured in hours, days, weeks, months or 
even years! The stau is not the only candidate for the NLSP. Other
generic possibilities include the lightest neutralino $\chi$, a sneutrino~\cite{sneutrino}, or even 
the lighter stop squark~\cite{stop}. In the case of minimal
supergravity (mSUGRA), one finds that a gravitino LSP combined with a stau NLSP
is as generic as the conventional neutralino LSP scenario, 
as seen by the light (yellow) shaded region in the left panel of Fig.~\ref{fig:GDM}.
All of these possibilities are constrained by astrophysics and cosmology, particularly
through limits on the decays of the metastable NLSP~\cite{CEFO}, and its bound states if it is
charged~\cite{Pospelov}. These effects may even improve the agreement
between cosmological nucleosynthesis calculations and the observed Lithium
abundances, as shown in the pink (darker) shaded region in the right panel of
Fig.~\ref{fig:GDM} for one particular non-mSUGRA scenario~\cite{CEFOS}. This scenario
could be probed only with a higher-energy LC~\cite{Cakir}.

\begin{figure}
\resizebox{1.0\columnwidth}{!}
{\includegraphics{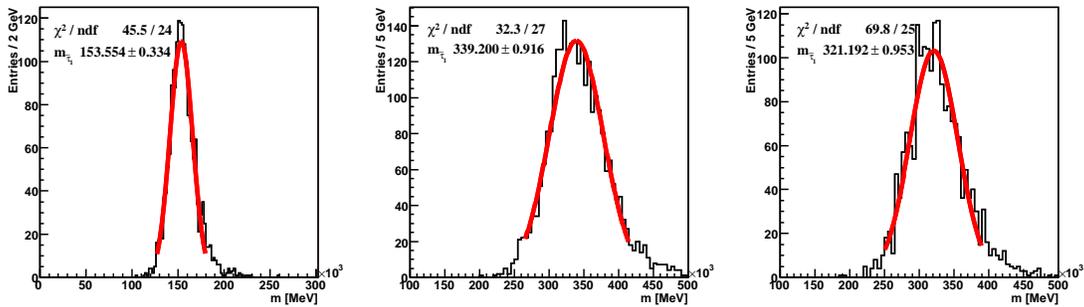}}
\caption{\label{fig:staumass}The mass of a metastable stau could be measured
quite accurately at the LHC~\protect\cite{ERO2}, as exemplified in three benchmark 
scenarios~\protect\cite{Bench3}. }
\end{figure}

\begin{figure}
\resizebox{1.0\columnwidth}{!}
{\includegraphics{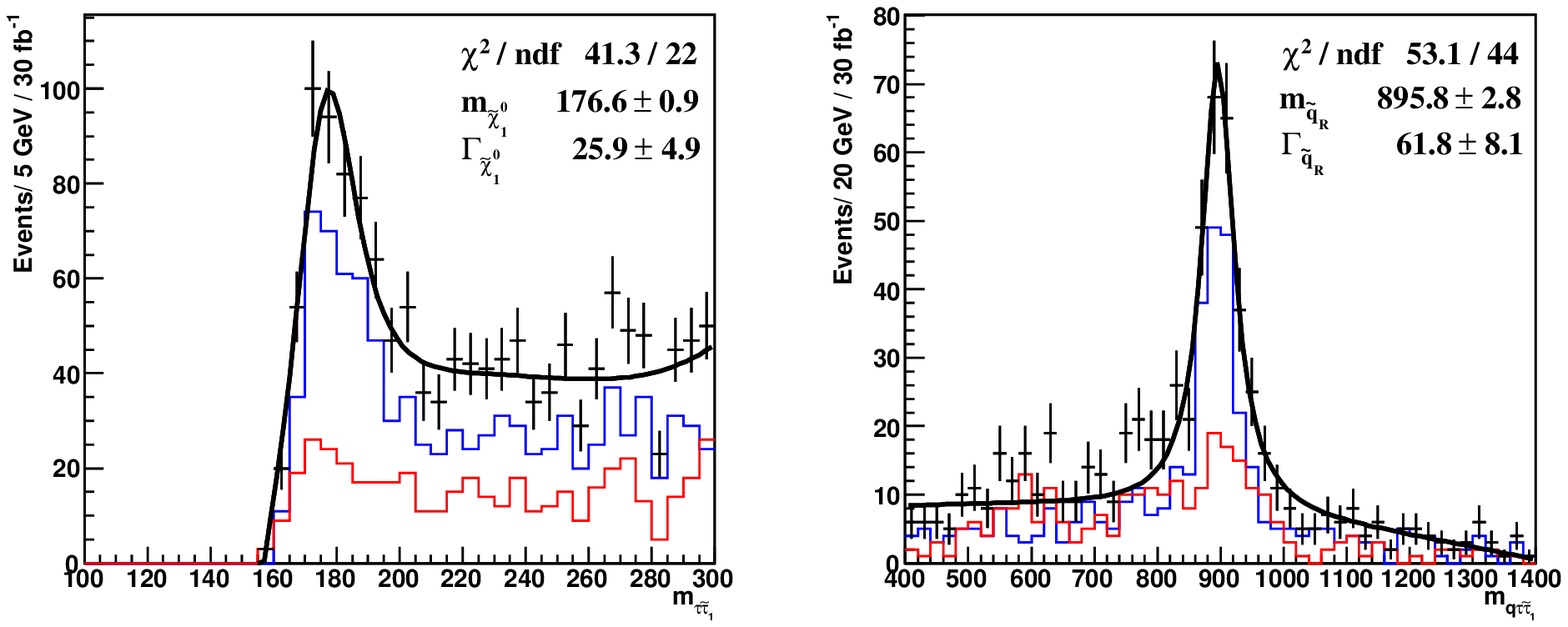}}
\caption{\label{fig:recontau}The reconstruction of heavier sparticles decaying into the ${\tilde \tau}$ in a scenario where it is the NLSP and the gravitino is the LSP~\protect\cite{ERO1}. }
\end{figure}

We recently examined how a stau NLSP scenario could be explored at the 
LHC~\cite{Bench3,ERO1,ERO2}.
We found that the normal experimental triggers on jets and energetic muons and 
electrons would  select the events containing staus~\cite{ERO1,ERO2}. 
It would then be possible to
identify the stau tracks in the events with quite high efficiency,
and measure the stau mass very
accurately via a combination of momentum and time-of-flight measurements,
as seen in Fig.~\ref{fig:staumass}.
The stau could be combined with jets in the event to reconstruct the masses of
heavier sparticles in the supersymmetric decay cascades, as in neutralino LSP
scenarios, as shown in Fig~\ref{fig:recontau}. 

Very slow-moving staus might be stopped either in the detector
material or in the rock surrounding the cavern~\cite{FS,Mihoko}. One could
use the muon system to locate the stau's impact point on the cavern wall with 
an uncertainty $\sim 1$~cm, and its impact angle with an accuracy $\sim 10^{-3}$.
One might then be able to bore into the cavern wall and remove a core from the
rock, which one could then store while waiting for the stau to decay.
Before boring, one would have to wait for a shut-down of the LHC,
since the radioactivity in the cavern is quite high while it is operating. However,
it is planned to stop the accelerator for a couple of days every month, so this
strategy might work if the lifetime exceeds about $10^6$~s!

\section{Extra Dimensions?}

Supersymmetry is certainly not the only possibility for new physics at the LHC.
Another possibility offered by string theory is that
there might be large extra dimensions. When string theory
was originally proposed as a `Theory of Everything', it was imagined that all
the extra dimensions would be curled up on length scales comparable to the
Planck length $\sim 10^{-33}$~cm. However, then it was realized that string
unification could be achieved more easily if one of these dimensions was
somewhat smaller than the GUT scale~\cite{HW}, and a number of scenarios with much
larger extra dimensions have been considered. For example, an extra
dimension of size $\sim 1$~TeV$^{-1}$ could help break supersymmetry~\cite{IA}
and/or the electroweak gauge symmetry, an extra dimension of micron size
could help rewrite the hierarchy problem~\cite{ADD}, and even infinite extra dimensions are
allowed if they are warped appropriately~\cite{RS}.

In many of these scenarios, there are potential signals to be found at the LHC,
such as Kaluza-Klein excitations of gravitons, or missing energy `leaking' into
an extra dimension. The most spectacular possibility would occur if gravity becomes
strong at the TeV energy scale, in which case microscopic black holes might be 
produced at the LHC. These would be very unstable, decaying rapidly via
Hawking radiation into multiple jets, leptons and photons, that would be easily
detectable~\cite{Webber} and distinguishable from supersymmetry and other
extra-dimensional scenarios~\cite{Smillie}.

\section{Summary}

The origin of mass is the most pressing problem in particle physics, and requires
a solution within the LHC energy range. Will it be a simple Higgs boson? 
and/or supersymmetry? The LHC will tell! there are many speculative ideas for 
other possible new physics beyond the Standard Model, such as
grand unification, strings, and extra dimensions. The LHC is also 
capable of probing many of these speculations including, as the last two
examples show, novel ideas that were undreamt of when the LHC and its
experiments were designed. We do not know what the LHC will find, but
we can be sure that its discoveries will set the agenda for possible future projects,
such as a linear $e^+ e^-$ collider.

\end{document}